\documentclass[aps,prd,preprintnumbers,showpacs,showkeys,nofootinbib,
superscriptaddress,fleqn,floatfix,tightenlines, 10pt]{revtex4-1}
\usepackage{amsmath,amsfonts,amssymb,amscd,amsxtra,amsthm}
\usepackage{graphicx}  
\usepackage{epstopdf}
\usepackage{dcolumn}  
\usepackage{bm}          
\usepackage{slashed}
\usepackage[utf8]{inputenc} 
\usepackage[normalem]{ulem} 
\usepackage[dvipsnames]{xcolor} 
\usepackage{array}
\usepackage{slashed}
\renewcommand\sout{\bgroup \color{red} \ULdepth=-.5ex \ULset}

\usepackage[caption = false]{subfig}
\usepackage{multirow}
\begin{document}  
\preprint{INHA-NTG-03/2017}
\title{Pion radiative weak decay from the instanton vacuum} 
\author{Sang-In Shim}
\email[E-mail: ]{ssimr426@gmail.com}
\affiliation{Department of Physics, Inha University, Incheon 22212,
Republic of Korea}
\affiliation{Research Center for Nuclear Physics (RCNP), Osaka 
University, Ibaraki, Osaka, 567-0047, Japan}
\author{Hyun-Chul Kim}
\email[E-mail: ]{hchkim@inha.ac.kr}
\affiliation{Department of Physics, Inha University, Incheon 22212,
Republic of Korea}
\affiliation{School of Physics, Korea Institute for Advanced Study 
  (KIAS),\\ Seoul 02455, Republic of Korea}
\date{\today}
\begin{abstract}
We investigate the vector and axial-vector form factors for the pion
radiative weak decays $\pi^+\to e^+\nu_e \gamma$ and $\pi^+\to
e^+\nu_e e^+ e^-$, based on the gauged effective chiral action from
the instanton vacuum in the large $N_c$ limit. The nonlocal
contributions, which arise from the gauging of the action, enhance the
vector form factor by about $20\,\%$, whereas the axial-vector form
factor is reduced by almost $30\,\%$. Both the results for the vector
and axial-vector form factors at the zero momentum transfer are in
good agreement with the experimental data. The dependence of the form 
factors on the momentum transfer is also studied. The slope parameters
are computed and compared with other works. 

\end{abstract}
\pacs{}
\keywords{}
\maketitle
\section{Introduction}
Pion radiative decay $\pi^+\to e^+ \nu_e \gamma$ provides rich
information on the structure of the pion. The decay amplitude for the
pion radiative decay consists of two part, i.e., the structure-dependent
(SD) part containing the vector and axial-vector form factors of the
pion and the inner Bremsstrahlung (IB)
part~\cite{Vaks:1958,Bludman:1960,Bryman:1982et,PDG}. The advantage of 
studying $\pi^+\to e^+ \nu_e \gamma$ decay over $\pi^+\to \mu^+
\nu_{\mu} \gamma$ is that the IB part is suppressed in the $\pi^+\to
e^+ \nu_e \gamma$ decay~\cite{Bryman:1982et,PDG}, whereas  
the corresponding SD part is enhanced due to the helicity. Thus,
the $\pi^+\to e^+ \nu_e \gamma$ decay allows one to get 
access to the structure of the pion experimentally. The vector  
form factor $F_V$ is related to the decay rate of the $\pi^0\to
\gamma\gamma$ decay~\cite{Vaks:1958,Gershtein:1955fb} by the vector
current conservation, so it was easier to find it using the lifetime
of $\pi^0$. On the other hand, it took many years to measure
unambiguously the axial-vector form factors~\cite{Depommier:1962jpa,
  Depommier:1963zza, Stetz:1977ge, Bay:1986kf, Piilonen:1986bv, 
  Egli:1986nk, Egli:1989vu, Bolotov:1990yq, Frlez:2003pe}. Some years
ago, PIBETA Collaboration~\cite{Bychkov:2008ws} conducted a precise
measurement of the pion weak form factors, reporting the values of the 
vector and axial-vector form factors respectively as
$F_V(0)=0.0258(17)$ and $F_A(0)=0.0117(17)$. The slope of the vector
form factor was also measured: $a_V=0.10(6)$, which is defined in the
parametrization of the vector form factor $  F_V(q^2) =
F_V(0)/(1-a_V q^2/m_\pi^2)$ near $q^2\approx 0$. There is yet the second
axial-vector form factor which comes into play when the photon is
virtual. The SINDRUM Collaboration~\cite{Egli:1989vu} reported the
first measurement of the decay $\pi^+\to e^+\nu_e e^+e^-$ in which the
off-mass-shell photon decays into $e^+e^-$, and yielded the second
axial-vector form factor to be $R_A(0)=0.059_{-0.008}^{+0.009}$. 

The vector and axial-vector form factors for the pion radiative decay
were studied in chiral perturbation theory~\cite{Holstein:1986uj,
  Bijnens:1996wm, Geng:2003mt, Mateu:2007tr, Unterdorfer:2008zz}, since the 
experimental data on the axial-vector form factor can be used to 
determine a part of the low-energy constants that encode information
on nonperturbative quark-gluon dynamics. These form factors have been
also investigated within various theoretical frameworks: For example,
quantum chromodynamics (QCD) sum rules~\cite{Nasrallah:1981rn}, the
nonlocal Nambu-Jona-Lasinio (NJL) model~\cite{Dumm:2010hh,
  GomezDumm:2012qh}, and in the light-front quark
model~\cite{Chen:2010ue}. Since the photon can be 
virtual, it is of interest to examine the dependence of the form
factors on the momentum transfer. Chiral perturbation theory predicts
very mild dependence on the momentum transfer in the range of $0\le
q^2\le 0.018\,\mathrm{GeV}^2$~\cite{Bijnens:1996wm}. On the other
hand, the results for the vector and axial-vector form factors from
the light-front quark model start to rise near $q^2=0$ and then fall
off drastically as $q^2$ increases~\cite{Chen:2010ue}. On the
contrary, the nonlocal NJL model~\cite{GomezDumm:2012qh} predicted
only the $q^2$ dependence of the vector form factor. The results
monotonically decrease as $q^2$ increases. Thus, it is of great
importance to investigate the weak form factors for the pion radiative
decay and compare them with those from other works. 

In the present work, we study the three weak form factors of the pion,
i.e., the vector form factor, the axial-vector form factor, and the
second axial-vector form factor, based on the gauged effective chiral
action (E$\chi$A) from the instanton vacuum~\cite{Diakonov:1985eg,
  Diakonov:2002fq, Musakhanov:2002xa, Kim:2004hd, Kim:2005jc,
Goeke:2007nc, Goeke:2007bj}. Since the
spontaneous breakdown of chiral symmetry is naturally realized from
the instanton vacuum, it provides a good framework to investigate
properties of the pion, i.e. of the pseudo-Nambu-Goldstone bosons. The
quark acquires the dynamical quark mass that is momentum-dependent
through the quark zero modes in the instanton background. Moreover,
there are only two parameters in this approach, namely, the average
instanton size $\bar{\rho} \approx 1/3$ fm and average interinstanton
distance $\bar{R} \approx 1$ fm. Since the average size of instantons
is considered as a normalization point equal to $\bar{\rho}^{-1}
\approx 0.6$ GeV, we can use the model for computing any observables
of hadrons and compare the results with those from other theoretical 
framework such as $\chi$PT and lattice QCD, in particular, when a
specific scale is involved. These values of the $\bar{\rho}$ and
$\bar{R}$ were determined many years ago
theoretically~\cite{Diakonov:1985eg,Diakonov:2002fq} as 
well as phenomenologically~\cite{Shuryak:1981ff,Schafer:1996wv}. They
were also confirmed by various lattice works~\cite{Chu:1994vi,
  Negele:1998ev, DeGrand:2001tm}. In Ref.~\cite{Cristoforetti:2006ar},
the QCD vacuum was simulated in the interacting instanton liquid model
and $\bar{\rho}\approx 0.32\,\mathrm{fm}$ and $\bar{R}\approx
0.76\,\mathrm{fm}$ were obtained with the finite current quark mass
taken into account. 

Since we consider the pion mass, we need to introduce the current
quark mass. Musakhanov~\cite{Musakhanov:1998wp,Musakhanov:2002vu}
improved the E$\chi$A derived by Diakonov and
Petrov~\cite{Diakonov:1985eg}, including the current quark mass. In
fact, this improvement plays an essential role in understanding the
QCD vacuum in the presence of the finite mass of the current quark. In
Ref.~\cite{Nam:2006ng}, it was shown that the improved E$\chi$A
properly described the dependence of the quark and 
gluon condensates on the current quark mass. Furthermore, the
nonlocality arising from the momentum-dependent dynamical quark mass
is known to bring out the breakdown of the Ward-Takahashi (WT)
identities, that is, the current nonconservation~\cite{Chretien:1954we,
  Pobylitsa:1989uq, Bowler:1994ir, Musakhanov:1996cv}. In
Refs.~\cite{Musakhanov:2002xa, Kim:2004hd}, the gauged E$\chi$A was
derived from the instanton vacuum, which satisfies the WT
identities. We will employ this action in the present work to
investigate the weak form factors for pion radiative decay.     

The structure of the present work is sketched as follows: In Section
II, we will define the three weak form factors of the pion, which will
be related to the transition matrix elements of the vector and
axial-vector currents. In Section III, we briefly explain the gauged
E$\chi$A. In Section IV, we derive the vector and axial-vector form
factors, using the gauged E$\chi$A. In Section V, we present the
numerical results of the three form factors and discuss them. The
final Section is devoted to summary and conclusion.  
\section{ Weak form factors of the $\pi^+ \to e^+ \nu \gamma$ decay } 
The SD part of the pion radiative decay amplitude consists of the
weak transition form factors of the pion, i.e., the vector form factor
$F_V(q^2)$, the axial-vector form factor $F_A(q^2)$, and the second
axial-vector form factors $R_A(q^2)$. They are defined in terms of the
transition matrix elements of the vector and axial-vector currents as
follows 
\begin{align}
\langle \gamma(k)|V_{12}^{\mu}(0)|\pi^+(p)\rangle
&=-\frac{e}{m_{\pi}}\epsilon_{\alpha}^{*} F_{V}(q^2) 
\epsilon^{\mu \alpha \rho \sigma} p_{\rho} k_{\sigma}, 
\label{eq:VectorMarix}
\\
\langle \gamma(k)|A_{12}^{\mu}(0)|\pi^+(p)\rangle
&=ie\epsilon_{\alpha}^{*}
\sqrt{2}f_\pi \left[
-g^{\mu \alpha}
+q^{\mu} (q^{\alpha}+p^{\alpha})\frac{ F_{\pi}(k^2)}{q^2-m_{\pi}^2}
\right]\cr
&\hspace{0.3cm}
+i\epsilon_{\alpha}^{*}\frac{e }{m_{\pi}}
\left[
F_A(q^2)\left(k^{\mu}q^{\alpha}-g^{\mu \alpha}q\cdot k\right)
+R_A(q^2)\left(k^{\mu}k^{\alpha}-g^{\mu\alpha}k^2\right)
\right],
\label{eq:AxialMarix}
\end{align}
where $|\pi^+(p)\rangle$ and $|\gamma (k) \rangle$ stand for the
initial pion and the final photon states, the transition vector and
axial-vector currents are defined 
respectively as 
\begin{align}
V_{12}^{\mu} = \bar{\psi} \gamma^\mu \frac{\tau_1-i \tau_2}{2}
  \psi,\;\;\;
A_{12}^{\mu} = \bar{\psi} \gamma^\mu\gamma_5 \frac{\tau_1-i \tau_2}{2}
  \psi, \label{eq:vacurrent}
\end{align}
consisting of the quark fields $\psi=(u,\,d)$, the Dirac matrices
$\gamma_\mu$ and $\gamma_5$, and the Pauli matrices $\tau_i$ in
isospin space. $p$ and $k$ denote respectively the momenta of the pion
and the photon, whereas $q$ is the momentum of the lepton pair. The
mass of the pion can be obtained from $p^2=m_\pi^2$ with the mass of
the pion $m_\pi= 139.57$ MeV. $F_V(q^2)$ and $F_A(q^2)$ are
the vector and axial-vector form factors of the pion respectively. The 
second axial-vector form factor, $R_A(q^2)$ contributes only when the 
outgoing photon is virtual($k^2\ne 0$). $F_{\pi}(k^2)$ is the
electromagnetic form factor which gives
$F_{\pi}(0)=1$. Electromagnetic charge radius $\langle r^2_\pi\rangle$  
and $F_{\pi}(k^2)$, were already calculated by one of the authors and 
his collaborator in this model~\cite{Nam:2007gf}. 

\section{Gauged effective chiral action in the presence of external fields}
Since we want to compute the weak form factors of pion radiative decay
in this work, we introduce all the relvant external fields in the
gauge-invariant manner, \textit{i.e.}, the electromagnetic field
$v^{\mathrm{em}}$, the vector fields $v$, and the axial-vector fields
$a$ 
\begin{align}
\mathcal{S}_{\mathrm{eff}}[v^{\mathrm{em}},v,a,\pi]
=-\mathrm{Sp}\ln \left[i\, \slashed{\mathcal{D}}+i\hat{m}
+i\sqrt{M(i\,\mathcal{D}^{L})}
  U^{\gamma_5}\sqrt{M(i\,\mathcal{D}^{R})} \right], 
\label{eq:GaugeEXA}
\end{align}
where the functional trace $\mathrm{Sp}$ runs over the
space-time, color, flavor, and spin spaces. The current quark mass
matrix $\hat{m}$ is written as $\mathrm{diag}(m_u,\,m_d)=\bar{m}\bm{1}
+ m_3\tau_3$ with $\bar{m}=(m_u+m_d)/2$ and
$m_3=(m_u-m_d)/2$. $\tau_3$ is the third component of the Pauli
matrix. Note that isospin symmetry is assumed, so $m_3=0$. The
covariant derivative $\mathcal{D}_\mu$ is defined as   
\begin{align}
i\,\mathcal{D}_{\mu} =
i\partial_{\mu} + e\hat{Q}\,v^{\mathrm{em}}_{\mu} 
+ \frac{\tau^a}{2} v^a_{\mu} + \gamma_5 \frac{\tau^a}{2} a^a_{\mu}
\end{align}
with the charge operator for the quark fields 
\begin{align}
\hat{Q}= \left(
\begin{array}{cc}
\frac{2}{3} 	& 0\\
0        	& -\frac{1}{3}
\end{array}
\right) = \frac{1}{6} + \frac{1}{2}\tau_{3}.
\end{align}
The left-handed and right-handed covariant derivatives in the
momentum-dependent dynamical quark mass $M(i\mathcal{D}_{L,R})$ are 
defined respectively as 
\begin{align}
i\,\mathcal{D}_{\mu}^L =
i\partial_{\mu} + e\hat{Q}\,v^{\mathrm{em}}_{\mu} 
+ \frac{\tau^a}{2} v^a_{\mu}  - \gamma_5 \frac{\tau^a}{2}
  a^a_{\mu},\;\;\;
i\,\mathcal{D}_{\mu}^R =
i\partial_{\mu} + e\hat{Q}\,v^{\mathrm{em}}_{\mu} 
+ \frac{\tau^a}{2} v^a_{\mu} + \gamma_5 \frac{\tau^a}{2} a^a_{\mu}.  
\label{eq:covLR}
\end{align}
The momentum-dependent quark mass with the covariant derivatives
ensures the gauge invariance of Eq.~(\ref{eq:GaugeEXA})in the presence
of the external fields. In fact, it was shown that the 
The nonlinear pseudo-Nambu-Goldstone boson field is
expressed as
\begin{equation}
\label{eq:NLGB}
U^{\gamma_5} = U(x) \frac{1+\gamma_5}{2} + U^\dagger (x)
\frac{1-\gamma_5}{2} = \exp \left(\frac{i
    \gamma_5}{f_\pi}\bm{\tau}\cdot\bm{\pi} 
\right),
\end{equation}
where $F_\pi$ is the pion decay constant. The pion fields are given as  
\begin{equation}
\label{eq:PGBF}
\bm{\tau} \cdot \bm{\pi} = \frac1{\sqrt{2}} \left( \begin{array}{cc}
\frac{1}{\sqrt{2}}\pi^0  & \pi^+  \\
\pi^- & -\frac{1}{\sqrt{2}}\pi^0
\end{array}\right).
\end{equation}
The momentum-dependent dynamical quark mass, which arises from the 
the quark-zero mode of the Dirac equation with the instanton fields,
is given by 
\begin{equation}
\label{eq:DQM}
M_f(k) = M_0 F^2(k) f(m_f),
\end{equation}
where $M_0$ is the constituent quark mass at zero quark virtuality,
and is determined by the saddle-point equation, resulting in about
$350$ MeV~\cite{Diakonov:1985eg,Diakonov:2002fq}. The
form factor $F(k)$ arises from the Fourier transform of the quark
zero-mode solution for the Dirac equation with the instanton and has
the following form:
\begin{equation}
\label{eq:ff_dqm}
F(k) = 2\tau\left[ I_0 (\tau) K_1(\tau) -I_1 (\tau)K_0(\tau)
  -\frac{1}{\tau}I_1(\tau)K_1(\tau) \right],
\end{equation}
where $\tau\equiv\frac{|k|\bar{\rho}}{2}$. $I_{0,1}$ and $K_{0,1}$
denote the modified Bessel functions. 
In addition to this original form, we also use the dipole-type
parametrization of $F(k)$ defined by 
\begin{equation}
\label{eq:ff_dipole}
F(k)= \frac{2 \Lambda^2}{2 \Lambda^2 + k^2}
\end{equation}
with $\Lambda=1/\bar{\rho}$. As mentioned in Introduction already, the
average size of the instanton $\bar{\rho}$ was determined either
phenomenonlogically~\cite{Shuryak:1981ff,Schafer:1996wv} or
theoretically~\cite{Diakonov:1985eg,Diakonov:2002fq}. In the large
$N_c$ limit, the value of $\bar{\rho}$ was determined to be
$\bar{\rho}\simeq 0.33$
fm~\cite{Diakonov:1985eg,Diakonov:2002fq}. When one considers the
$1/N_c$ meson-loop corrections, $\bar{\rho}$ is modified to be
$\bar{\rho}\simeq 0.35$ fm~\cite{Kim:2004hd, Kim:2005jc,
Goeke:2007nc, Goeke:2007bj}. Lattice QCD yields similar results
$\bar{\rho}=(0.32-0.36)$ fm~\cite{Chu:1994vi, Negele:1998ev,
  DeGrand:2001tm, Cristoforetti:2006ar}. Since we compute in this work
the weak form factors of pion radiative decay in the large $N_c$
limit, we will take $\bar{\rho}=0.33$ fm or $\Lambda = 600$ MeV.  
We will compare the results obtained by
using the both form factors. The presence of the current quark mass
also affects the dynamical one, which was studied in
Refs.~\cite{Musakhanov:1998wp,Musakhanov:2002vu} in detail.  
The additional factor $f(m_f)$ describes the $m_f$ dependence of the
dynamical quark mass, which is defined
as~\cite{Pobylitsa:1989uq,Musakhanov:2001pc}   
\begin{align}
f(m_f) =  \sqrt{1+\frac{m_f^2}{d^2}} -  \frac{m_f}{d}.
\end{align}
This $m_f$-dependent dynamical quark mass yields the gluon condensate
that does not depend on $m_f$. Pobylitsa considered the sum of
all planar diagrams, expanding the quark propagator in the instanton
background in the large $N_c$ limit~\cite{Pobylitsa:1989uq}. Taking
the limit of $N/(VN_c)\to 0$ leads to $f(m_f)$. The parameter
$d$ is given as $198$ MeV. The $m_f$-dependent dynamical quark mass
also explains a correct hierarchy of the chiral 
condensates: $\langle\bar{u}u\rangle\approx\langle\bar{d}d\rangle>
\langle\bar{s}s\rangle$~\cite{Nam:2006ng}.
\section{Pion weak form factors} 
\begin{figure}[htp]
\includegraphics[width=15cm]{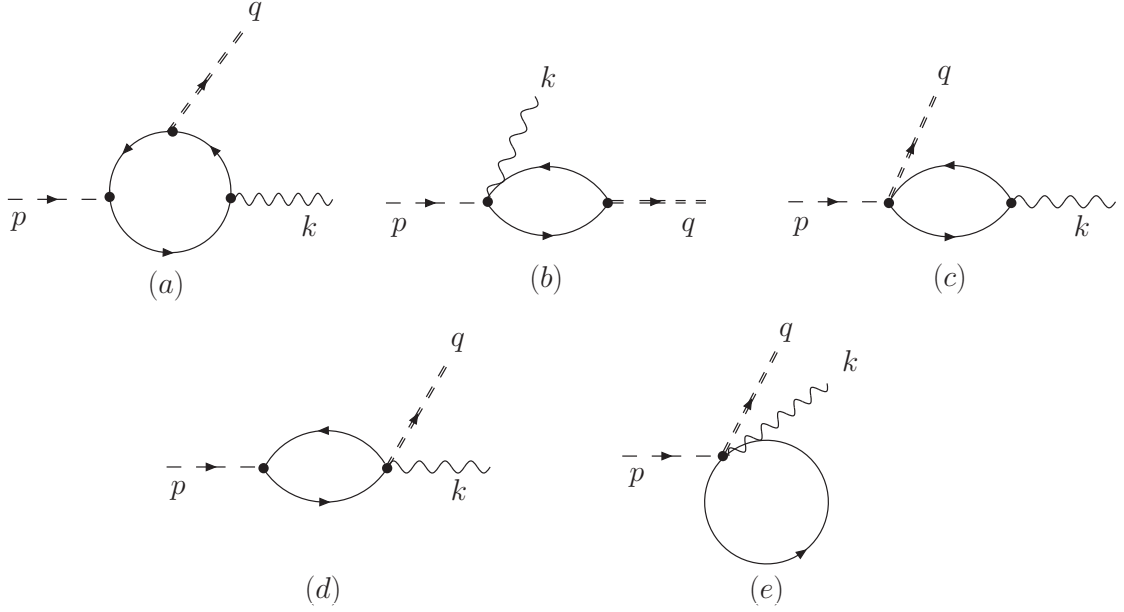}
\caption{The Feynman diagrams for the pion weak form factors. The 
dashed line depicts the pion, the dashed double line and
the wavy line describe the vector (axial-vector) field and the photon
field, respectively. Diagram (a) contains both the local and nonlocal
contributions, whereas diagrams (b)-(e) arise from the nonlocal
interaction due to the momentum-dependent dynamical quark
mass.}
\label{fig:1}
\end{figure}

The matrix elements of the vector and axial vector currents in
Eq.(\ref{eq:AxialMarix}) are related to the three-point correlation
function 
\begin{align}
\langle\gamma (k)|\mathcal{W}_\mu^{a}(0) |\pi^b(p)\rangle
=\epsilon^*_\alpha\int d^4x e^{-ik\cdot x} \int d^4y e^{ip\cdot y}
 \mathcal{G}_{\alpha \rho}^{-1}(k)\mathcal{G}_{\pi}^{-1}(p)\langle
  0|\{V^{\mathrm{em}}_{\rho}(x)  \mathcal{W}_{\mu}^{a} (0)
  P^b(y)\}|0\rangle,  
\label{eq:matrixel}
\end{align}
where $\mathcal{W}_\mu^{a}$ expresses generically either the vector
current or the axial-vector current defined in
Eq.(\ref{eq:vacurrent}). The operators in the correlation function
represent the electromagnetic current, vector (axial vector) current,
and pion-field operators, respectively. $\mathcal{G}_{\alpha\rho}
(k)$, $\mathcal{G}_{\pi} (p)$ stand for the propagators of the photon
and the pion, respectively.  
Then, the matrix element~(\ref{eq:matrixel}) can be directly derived
from the gauged effective chiral action given in Eq.(\ref{eq:GaugeEXA}) 
\begin{align}
\langle\gamma (k)|\mathcal{W}_\mu^{a}(0) |\pi^b(p)\rangle
=\epsilon^*_\alpha\int d^4x e^{-ik\cdot x} \int
  d^4y e^{ip\cdot y}  \left. \frac{\delta^3
  \mathcal{S}_{\mathrm{eff}}[ v^{\mathrm{em}}, w, 
  \pi ] }{\delta v_\alpha^{\mathrm{em}}(x) \delta w_\mu^a(0)\delta
  \pi^{b}(y)}   \right|_{v^{\mathrm{em}},w,\pi=0}. 
\label{eq:corr}
\end{align}
The three-point correlation function in Eq.(\ref{eq:corr}) consists of
five Feynman diagrams drawn in Fig.~\ref{fig:1}. In the case of the
vector form factor, only diagram (a) contributes to it, whereas all
other diagrams vanish because of the trace over spin space. On the
other hand, all the diagrams contribute to the axial-vector form
factors.      
Note that diagram (a) contains the contributions from both the local
and nonlocal terms, while all other diagrams arise only from the
nonlocal terms on account of the momentum-dependent dynamical quark
mass. 
\subsection{Vector form factor}
We first deal with the vector form factor of the pion. Having computed 
Eq.(\ref{eq:corr}) explicitly, we obtain the matrix element of the
vector current ($\mathcal{W}=V$)  
\begin{align}
\langle\gamma(k)|V_{\mu}^{12}|\pi^+(p)\rangle
&= -i\frac{4\sqrt{2}e N_c}{3f_\pi}
\epsilon^{*}_{\alpha} \int
\frac{d^4 l}{(2\pi)^{4}}
\frac{
\sqrt{M(k_a) M(k_b)}
}{
\mathcal{D}_{a}\mathcal{D}_{b}\mathcal{D}_{c}
}
\left[\varepsilon_{\mu\alpha\rho\sigma}
\left(\bar{M}_a k_{b\rho} k_{c\sigma} + \bar{M}_b k_{c\rho} k_{a\sigma}
+ \bar{M}_c k_{a\rho} k_{b\sigma} \right) \right.
\cr
&\hspace{0.4cm} 
- \varepsilon_{\mu\beta\rho\sigma} k_{a\beta} k_{b\rho} k_{c\sigma}
\left( \sqrt{M(k_b)}\sqrt{M_{\alpha}(k_b)} 
+\sqrt{M(k_c)}\sqrt{M_{\alpha}(k_c)} \right)
\cr
&\hspace{0.4cm}
\left.+ \varepsilon_{\alpha\beta\rho\sigma} k_{a\beta} k_{b\rho} k_{c\sigma} 
\left(\sqrt{M(k_a)}\sqrt{M_{\mu}(k_a)} 
+ \sqrt{M(k_c)}\sqrt{M_{\mu}(k_c)} \right)
\right],
\label{eq:VecMatElmt}
\end{align}
where $N_c$ denotes the number of colors. $\bar{M}_i$ is the sum of
the dynamical and current quark masses $\bar{M}_i=m+M(k_i)$. The
momenta $k_i$ are defined as 
$k_a=l+\frac{q}{2}+\frac{k}{2}$, $k_b=l-\frac{q}{2}-\frac{k}{2}$,
$k_c=l-\frac{q}{2}+\frac{k}{2}$, and $q=p-k$. $\mathcal{D}_i$ are
given as $\mathcal{D}_{i}=(k_i^2+\bar{M}_i^2)$. $\sqrt{M_\mu(k_i)}$
represents $\sqrt{M_\mu(k_i)} = \partial \sqrt{M(k_i)}/\partial
k_{i\mu}$. Equation~(\ref{eq:VecMatElmt}) corresponds to diagram (a)
in Fig.~\ref{fig:1} and there is no contribution from diagrams (b)-(e)
in the case of the vector form factor, as mentioned previously. 
Considering the transverse relation $\epsilon^{*}\cdot p=\epsilon
\cdot p=0$, we can extract the vector form factors, comparing
Eq.(\ref{eq:VectorMarix}) with Eq.(\ref{eq:VecMatElmt}). Thus,  
the pion vector form factor is obtained finally as 
\begin{align}
F_V(Q^2) = F_V^{\mathrm{local}} (Q^2) + F_V^{\mathrm{NL}}(Q^2),   
\end{align}
where $F_V^{\mathrm{local}} (Q^2)$ and $F_V^{\mathrm{NL}}(Q^2)$
stand for the local and nonlocal contributions 
\begin{align}
F_V^{\mathrm{local}}(Q^2)
&= \frac{4\sqrt{2} N_c M_{\pi}}{3f_\pi (p\cdot k)^2}
\int \frac{d^4 l}{(2\pi)^{4}} 
\frac{\sqrt{M(k_a)M(k_b)}}{D_{a} D_{b} D_{c}} p_{\mu} k_{\nu}\bigg[
\bar{M}_a ( k_{b\mu} k_{c\nu} - k_{c\mu} k_{b\nu} ) 
\cr
& \hspace{5.4cm}+ \bar{M}_b ( k_{c\mu} k_{a\nu} - k_{a\mu} k_{c\nu} ) 
+ \bar{M}_c ( k_{a\mu} k_{b\nu} - k_{b\mu} k_{a\nu} )  
\Big],
\label{eq:localv}
\\
F_V^{\mathrm{NL}}(Q^2)
&=\frac{4\sqrt{2} N_c M_{\pi}}{3f_\pi (p\cdot k)^2}
\int\frac{d^4 l}{(2\pi)^{4}}
\frac{\sqrt{M(k_a)M(k_b)}}{\mathcal{D}_{a}\mathcal{D}_{b}\mathcal{D}_{c}} 
\bigg[-\Big(M'(k_b)+M'(k_c)\Big)(p\cdot k)^2
(\epsilon^{*}\cdot l)(\epsilon\cdot l) 
\cr
&\hspace{5.1cm}
+\Big( M'(k_a) + M'(k_c) \Big) 
(\varepsilon_{\mu\gamma\delta\lambda}l_{\mu}\epsilon_{\gamma}
p_{\delta}k_{\lambda})
(\varepsilon_{\alpha\beta\rho\sigma}\epsilon^{*}_{\alpha}
k_{a\beta}k_{b\rho}k_{c\sigma})
\bigg],
\label{eq:nonlocalv}
\end{align}
where $M'(k_i)$ is the derivative of the dynamical quark mass with
respective to the squared momentum $M'(k_i)=\partial M'(k_i)/ \partial
k_i^2 $. The momentum transfer $Q^2$ is defined to be positive
definite, \textit{i.e.}, $Q^2=-q^2$. 

In fact, one can easily see from Eq.~(\ref{eq:nonlocalv}) that the
terms with $M'(k_i)$ are derived from the expansion of the dynamical
quark mass with respect to the covariant detivative given in
Eq.~(\ref{eq:covLR}). Thus, those terms with $M'(k_i)$ are the
essential part in obtaining the vector and axial-vector form factors
with the corresponding gauge invariance preserved. 
If the dynamical quark mass is taken to be independent of the
quark momentum, then $M'(k_i)$ is equal to zero. It indicates that the
nonlocal contributions to the vector form factor vanish such that the
results are the same as those derived from the local chiral quark
model ($\chi$QM). However, one has to introduce the regularization to
tame the divergence arising from the quark loop in the local
$\chi$QM. In this sense, the momentum-dependent dynamical quark mass
plays also a role of a certain regularization. 

\subsection{Axial-vector form factors}
 The transition matrix element of the axial-vector current
 ($\mathcal{W}=A$) given in Eq.(\ref{eq:AxialMarix}) is 
 obtained as 
\begin{align}
\langle\gamma(k)|A_{\mu}^{12}|\pi^+(p)\rangle
=-i\frac{4\sqrt{2}eN_c}{f_\pi}
\epsilon^{*}_{\alpha}\int\frac{d^4 l}{(2\pi)^4}
\sum^{e}_{i=a}\mathcal{F}^{(i)}_{\mu\alpha},
\label{eq:axialVFF1}
\end{align}
where $\mathcal{F}^{(i)}_{\mu\alpha}$ corresponds to diagram ($i$),
which can be explicitly expressed as   
\begin{align}
\mathcal{F}^{(a)}_{\mu\alpha}
&=\frac{\sqrt{M(k_a)
  M(k_b)}}{\mathcal{D}_{a}\mathcal{D}_{b}\mathcal{D}_{c}}\left
  [\delta_{\mu\alpha}  \left\{ 
\bar{M}_a k_{b}\cdot k_{c} 
-\bar{M}_b k_{c}\cdot k_{a} +\bar{M}_c k_{a}\cdot k_{b}+\bar{M}_{abc}
\right\}\right. \cr
&+\left\{
-\bar{M}_a(k_{b\mu}k_{c\alpha} + k_{c\mu}k_{b\alpha})
+\bar{M}_b(k_{a\mu}k_{c\alpha}+k_{c\mu}k_{a\alpha})
+\bar{M}_c(k_{a\mu}k_{b\alpha}-k_{b\mu}k_{a\alpha})
\right\}\cr
&-\left(
M'(k_a)k_{a\mu}-M'(k_c)k_{c\mu}\right)\left\{
 (k_{b}\cdot k_{c}+\bar{M}_{bc})k_{a\alpha}
-(k_{c}\cdot k_{a}+\bar{M}_{ca})k_{b\alpha} 
-(k_{a}\cdot k_{b}+\bar{M}_{ab})k_{c\alpha}\right\}\cr
& + \left(
M'(k_b)k_{b\alpha}+M'(k_c)k_{c\alpha}\right)\left\{
-(k_{b}\cdot k_{c}-\bar{M}_{bc})k_{a\mu}
+(k_{c}\cdot k_{a}-\bar{M}_{ca})k_{b\mu}
-(k_{a}\cdot k_{b}+\bar{M}_{ab})k_{c\mu}\right\} \cr
& -\left. \left(M'(k_a)k_{a\mu}-M'(k_c)k_{c\mu}\right)
\left(M'(k_b)k_{b\alpha}+M'(k_c)k_{c\alpha}\right)\left\{
\bar{M}_a k_{b}\cdot k_{c}
-\bar{M}_b k_{c}\cdot k_{a}
-\bar{M}_c k_{a}\cdot k_{b}-\bar{M}_{abc}
\right\}\right], \cr
\mathcal{F}^{(b)}_{\mu\alpha}
&=\frac{\sqrt{M(k_a) M(k_b)}}{\mathcal{D}_{a}\mathcal{D}_{c}}
\frac{ M'(k_b)k_{b\alpha}}{M(k_b)}
\left[-\left\{\bar{M}_c+M'(k_a)(k_{a}\cdot k_{c}+\bar{M}_{ac})\right\}
  k_{a\mu}
+\left\{\bar{M}_a+M'(k_c)(k_{a}\cdot
  k_{c}+\bar{M}_{ac})\right\} k_{c\mu}\right], \cr 
\mathcal{F}^{(c)}_{\mu\alpha}
&=\frac{\sqrt{M(k_a)M(k_b)}}{\mathcal{D}_{b}\mathcal{D}_{c}}
\frac{M'(k_a) k_{a\mu}}{M(k_a)}\left[
-\left\{\bar{M}_c-M'(k_b)(k_{b}\cdot k_{c}-\bar{M}_{bc})\right\}
  k_{b\alpha} 
-\left\{\bar{M}_b-M'(k_c)(k_{b}\cdot k_{c}-\bar{M}_{bc})\right\}
  k_{c\alpha} \right], \cr
\mathcal{F}^{(d)}_{\mu\alpha}
&=\frac{\sqrt{M(k_a)M(k_b)}}{\mathcal{D}_{a}\mathcal{D}_{b}}
M(k_c)\left(\frac{M'(k_c)}{M(k_c)}\right)^2
\left(k_{a} \cdot k_{b} +\bar{M}_{ab}\right)
k_{c\mu}k_{c\alpha}, \cr
\mathcal{F}^{(e)}_{\mu\alpha}
&=\sqrt{M(k_a)M(k_b)}\frac{1}{\mathcal{D}_c}
\bar{M}_c\left(\frac{M'(k_a)M'(k_b)}{M(k_a)M(k_b)}\right)
k_{a\mu}k_{b \alpha}.
\label{eq:fas}
\end{align}
Here, $\bar{M}_{ij}=\bar{M}_a \bar{M}_b$ and $\bar{M}_{ijk}= \bar{M}_a
\bar{M}_b \bar{M}_c$. 

In order to pick up the axial-vector form factors from
Eq.(\ref{eq:AxialMarix}), it is convenient to introduce an arbitrary
vector $\xi^\perp_{\mu}$ that satisfies the following properties,
$\xi^\perp \cdot \xi^\perp =0$, $\xi^\perp \cdot q=0$, and $\xi^\perp \cdot
k\ne0$. Then, the axial-vector form factor $F_A(Q^2)$ and the second
axial-vector form factor $R_A(Q^2)$ can be derived as 
\begin{align}
F_{A}(Q^2)
&=-\frac{4\sqrt{2} N_c m_\pi}{f_\pi (q \cdot k)}
\int \frac{d^4 l}{(2 \pi)^4}
\sum^{e}_{i=a}\mathcal{F}^{(i)}_{\mu\alpha}
\left[ \epsilon_\mu \epsilon^{*}_{\alpha}
-\frac{\xi^\perp_{\mu} k_{\alpha}}{\xi^\perp \cdot k}
\right],
\\
R_{A}(Q^2)
&=\frac{4\sqrt{2} N_c m_\pi}{f_\pi (\xi^\perp \cdot k)^2}
\int \frac{d^4 l}{(2 \pi)^4}
\sum^{e}_{i=a}\mathcal{F}^{(i)}_{\mu\alpha}
\xi^\perp_\mu \xi^\perp_\alpha.
\end{align}

As in the vector form factor, the local contribution to the
axial-vector form factors comes from the first and second terms of 
$\mathcal{F}_{\mu \alpha}^{(a)}$ in Eq.~(\ref{eq:fas}). 

\section{Results and discussion}
We are now in a position to discuss the numerical results for the weak
form factors of the pion radiative decay. Since the present framework
is fully relativistic, the Breit-momentum frame will be
used. There are no adjustable parameters in the present work. We will
take the original values $M_0=350$ MeV and
$\bar{\rho}=0.33$ fm from
Refs~\cite{Diakonov:1985eg,Diakonov:2002fq}.  The pion decay constant
$f_\pi$ can be computed within the model and is obtained to be
$f_\pi=93$ MeV. 

\begin{figure}[htp]
\captionsetup[subfigure]{labelformat=empty}
\subfloat[]{\includegraphics[width = 2.4in]{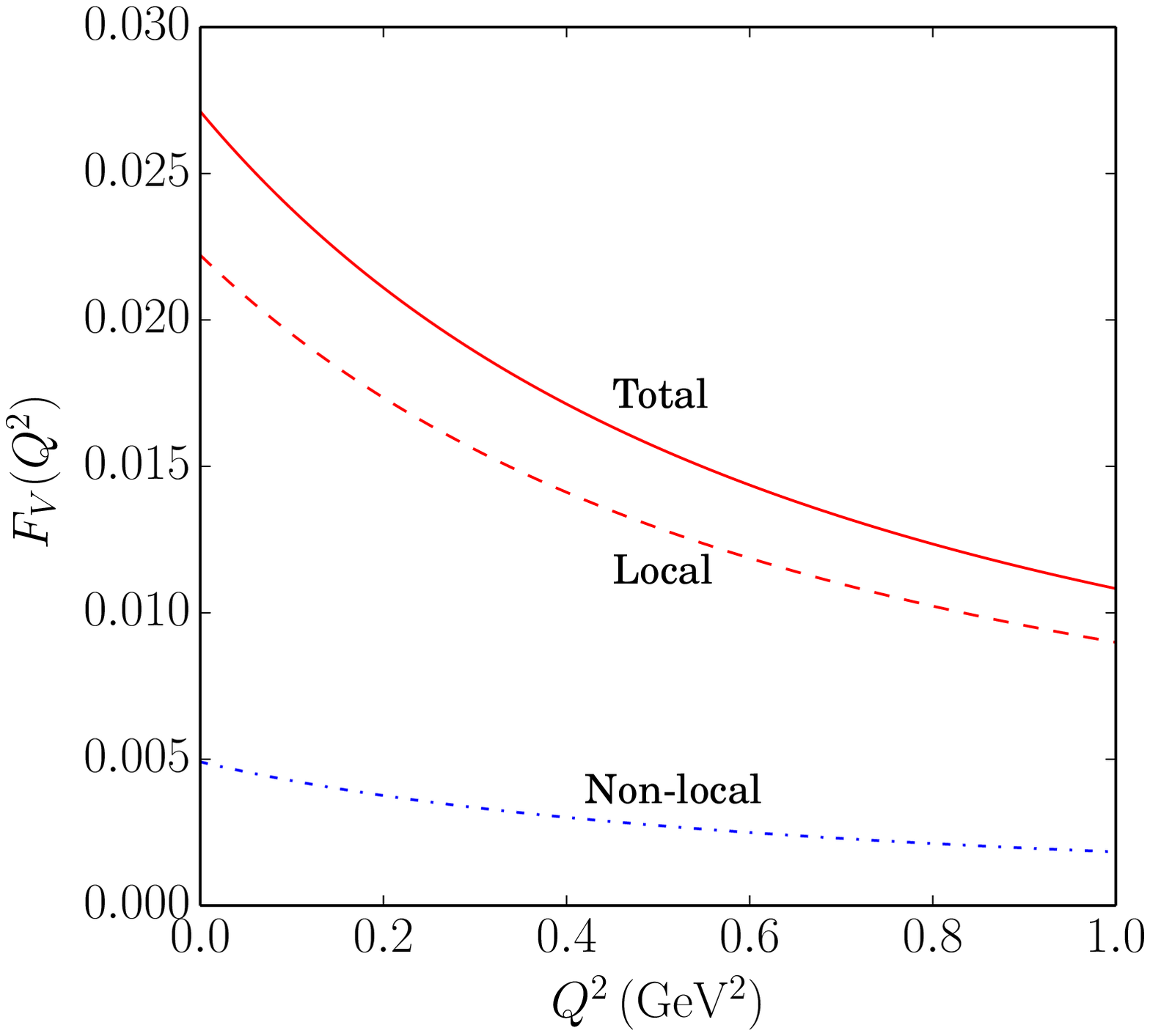}} 
\subfloat[]{\includegraphics[width = 2.4in]{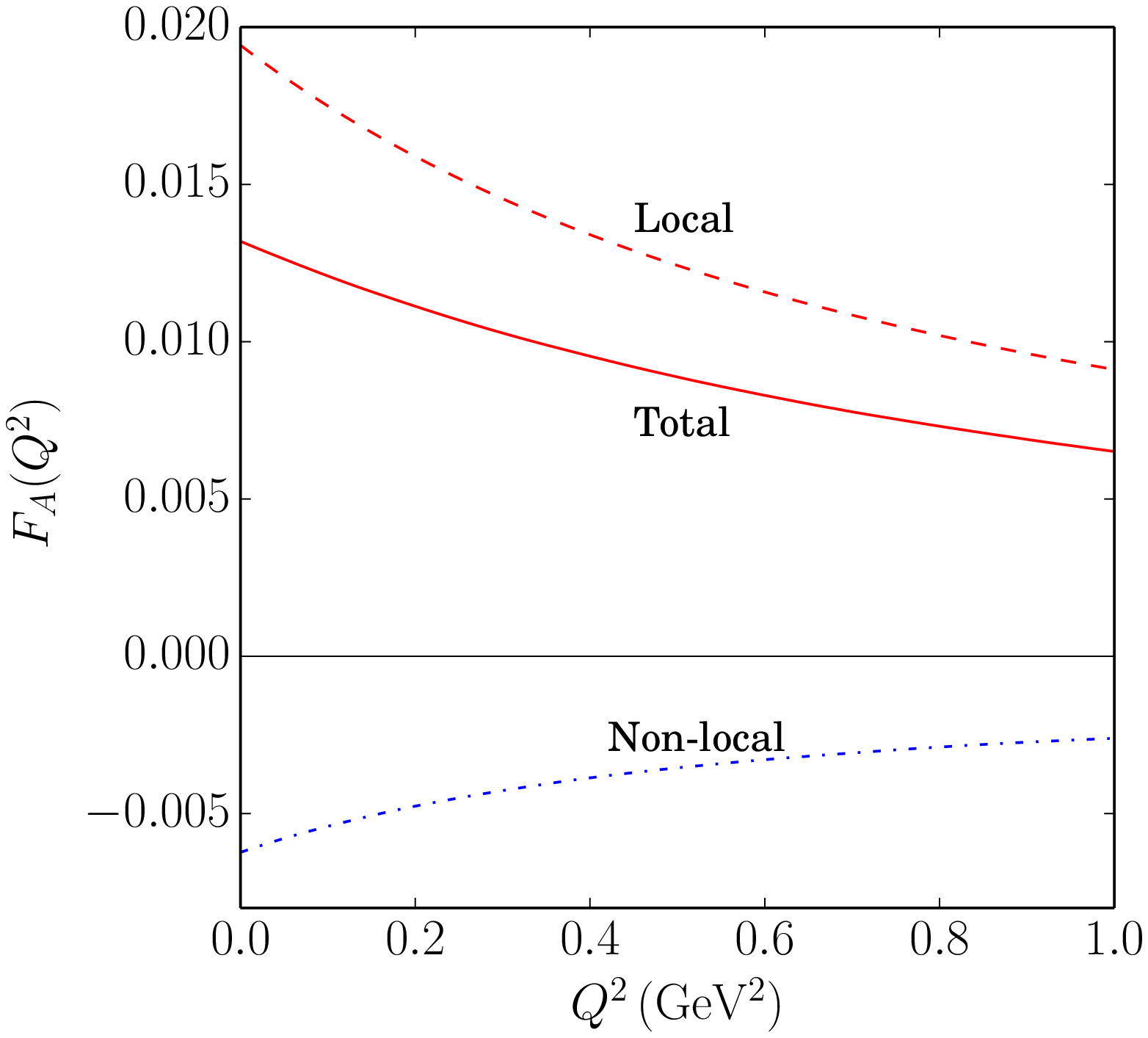}}
\subfloat[]{\includegraphics[width = 2.4in]{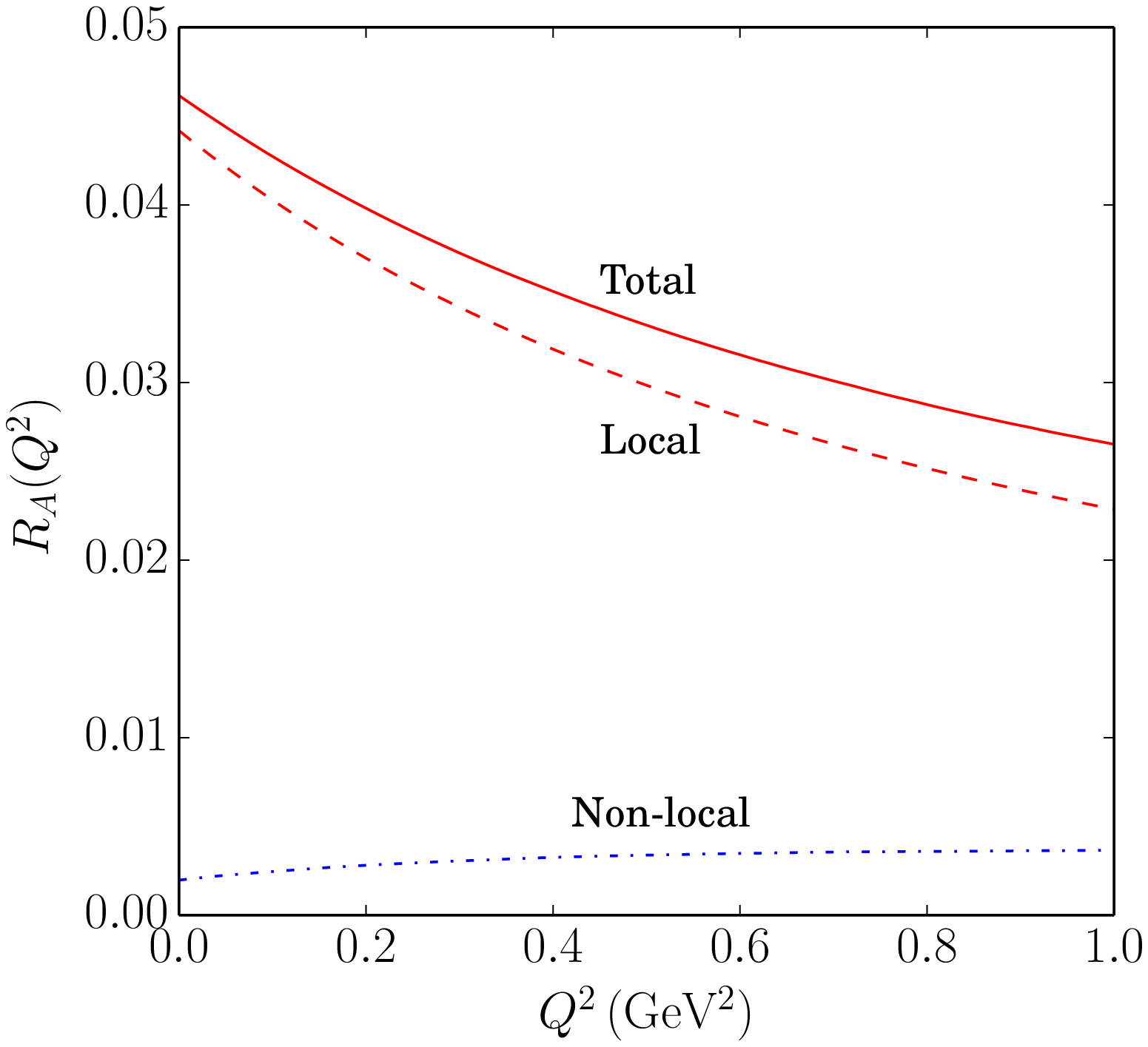}}
\caption{
The form factors $F_V(Q^2)$, $F_A(Q^2)$, and $R_A(Q^2)$ for the
radiative pion decay as functions of $Q^2$. 
In the left panel the vector form factor is depicted. The middle 
and the right panels correspond to the axial-vector form factor and the second
axial-vector form factor, respectively.  
The momentum-dependent quark mass defined in Eq.~(\ref{eq:DQM}) with
Eq.~(\ref{eq:ff_dqm}) was used. The dashed curve presents the local
contribution whereas the dot-dashed one draws the nonlocal
contribution. The solid curve depicts the total result.
  }
\label{fig:2}
\end{figure}

Figure~\ref{fig:2} draws the results of the pion form factors for pion
radiative weak decay. In general, the form factors decrease
monotonically, as $Q^2$ decreases.    
As discussed in the previous section, it is essential to consider the
nonlocal contribution to preserve the corresponding gauge  
invariances, since the electromagnetic and vector currents should be
conserved. As shown in the left panel of Fig.~\ref{fig:2}, the
nonlocal part contributes to the pion vector form factor by almost
about 20~\%.     

The results for the axial-vector form factor is
depicted in the middle panel of Fig.~\ref{fig:2} as a function of
$Q^2$. Note that, however, the nonlocal contribution behaves very
differently from the case of the vector form factor. In fact, it turns  
out negative, so that the final result for the form factor is reduced
by about 30~\%, which implies that it is indeed crucial to consider
the nonlocal part in computing the axial-vector form factor. As will
be discussed later, it is very important to take into account the
nonlocal part to describe the experimental data at $Q^2=0$.

The second axial-vector form factor $R_A(Q^2)$ comes into play, when
the momentum of the photon is virtual. That is, one can get access to
it by $\pi^+\to e^+ \nu_e e^+ e^-$ decay in which the virtual photon
is annihilated into $e^+$ and $e^-$. As shown in the right panel of
Fig.~\ref{fig:2}, the $Q^2$ dependence of $R_A(Q^2)$ is similar to
$F_A(Q^2)$. However, the nonlocal contribution is relatively small and
positive. Moreover, it starts to increase as $Q^2$ increases, which
makes $Q^2$ dependence slightly milder than those of the vector and
axial-vector form factors. Note that the nonlocal contribution becomes
saturated as $Q^2$ further increases.

\begin{figure}[htp]
\captionsetup[subfigure]{labelformat=empty}
\subfloat[]{\includegraphics[width = 2.4in]{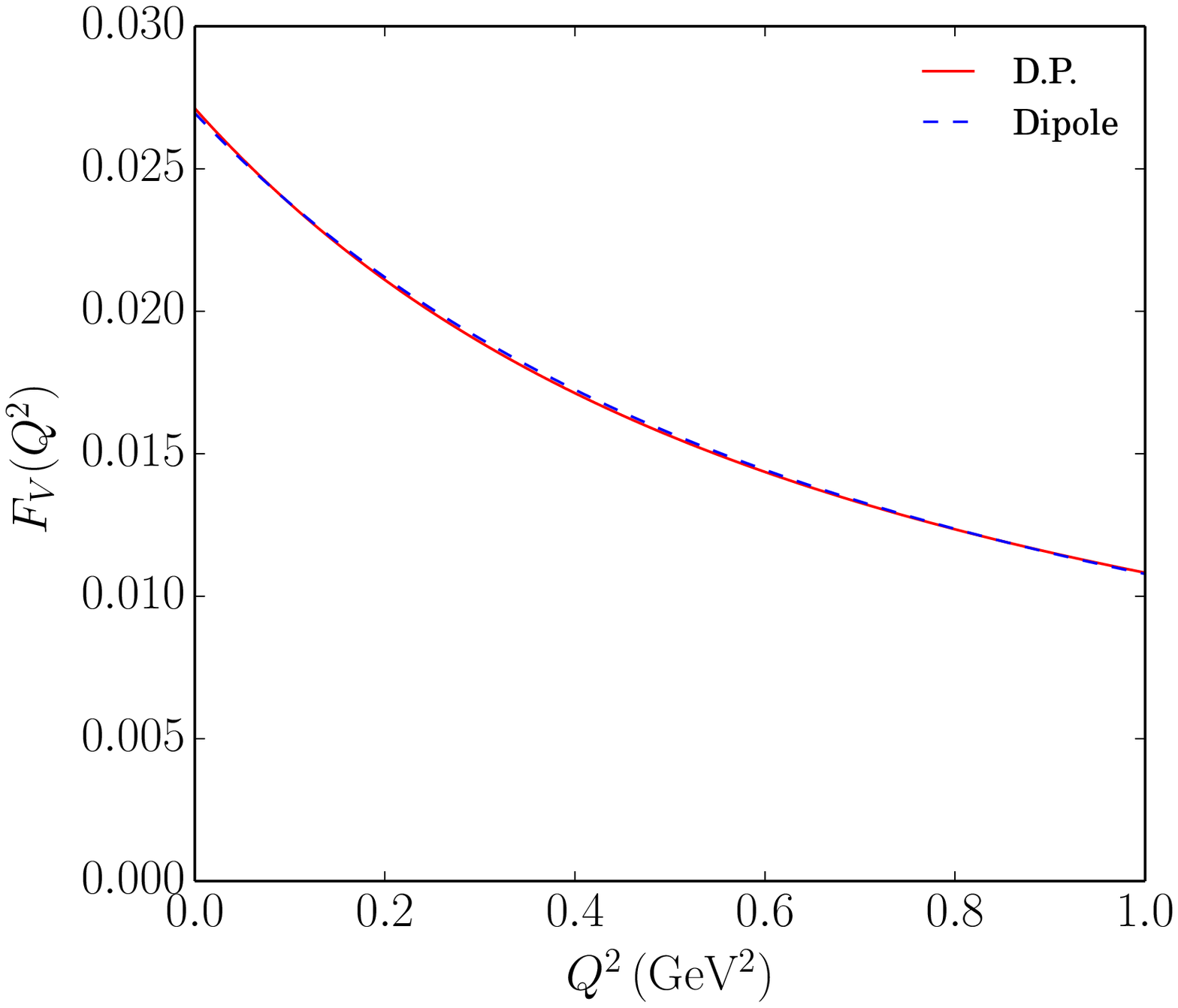}} 
\subfloat[]{\includegraphics[width = 2.4in]{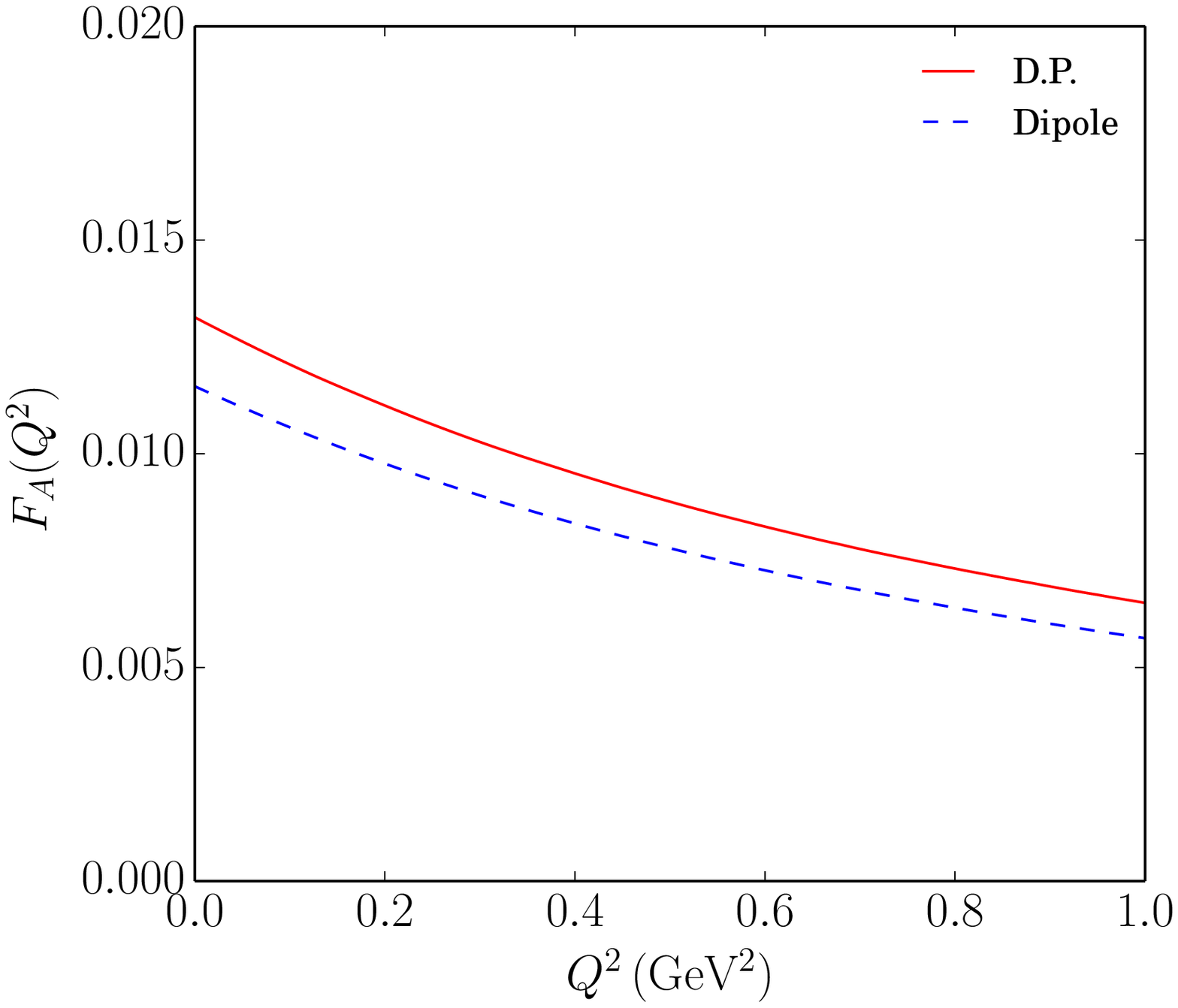}}
\subfloat[]{\includegraphics[width = 2.4in]{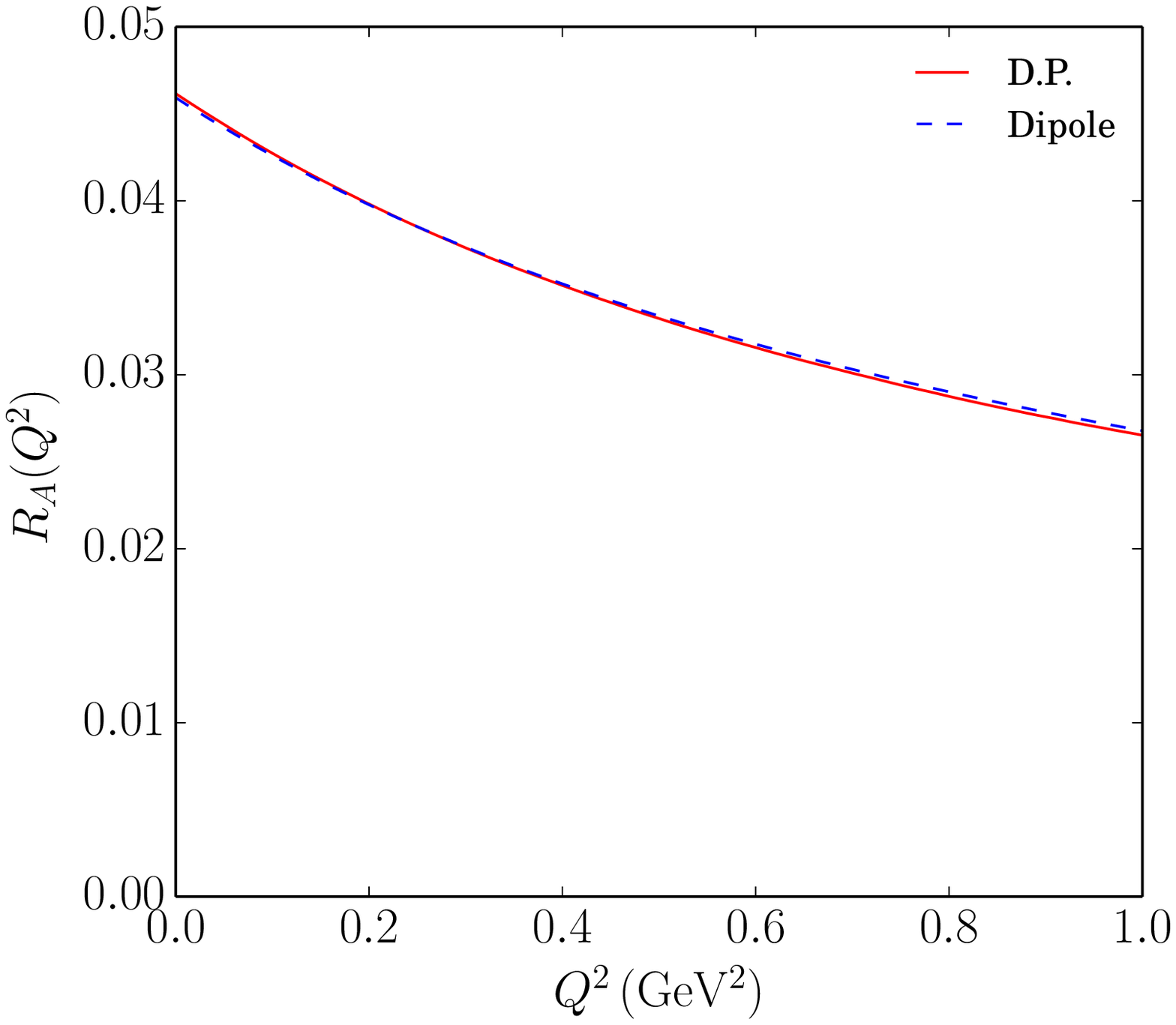}}
\caption{Comparisons of the form factors, $F_V(Q^2)$, $F_A(Q^2)$
    and $R_A(Q^2)$ with two different types of the momentum-dependent
    quark mass used. The solid curves draw the form factors derived
    by using the quark mass from the instanton vacuum given in
    Eq.~(\ref{eq:ff_dqm}) (D.P.), whereas the dashed ones depict that by
    the dipole-type mass given in Eq.~(\ref{eq:ff_dipole}) (Dipole).}  
\label{fig:3}
\end{figure}
In Fig.~\ref{fig:3}, we compare two different results of the pion weak 
form factors, employing the two different forms of the dynamical quark
mass given in Eq.~(\ref{eq:ff_dqm}) and Eq.~(\ref{eq:ff_dipole}),
respectively. The dynamical quark mass with the dipole-type
parametrization yields almost the same results for the vector and
second axial-vector form factors. On the other hand, it
gives a smaller result for the axial-vector form factor by around
$12\,\%$ in comparison with that from the instanton vacuum. 

\begin{table}[htp]
\centering
\caption {Comparison of the present results with those from other
  works. ``D.P.'' stands for the results derive from the instanton vacuum,
whereas ``Dipole'' denotes those obtained from the dipole-type
parametrization of the dynamical quark mass.}
\label{tab:1} 
\begin{tabular}{m{2.5cm}m{2.2cm}m{3.0cm}m{2.5cm}m{3.3cm}|m{2cm}m{1cm}}
\hline\hline
&\multirow{2}{*}{NJL~\cite{Courtoy:2007vy}} 
& \multirow{2}{*}{NL NJL(A)~\cite{GomezDumm:2012qh}} 
& \multirow{2}{*}{$\chi$PT~\cite{Unterdorfer:2008zz}} 
& \multirow{2}{*}{Experimental data} 
& \multicolumn{2}{l}{ $\hspace{0.5cm}$ Present work} \\ \cline{6-7} & & & & 
& $\hspace{0.1cm}$ D.P.  & Dipole  \\ \hline
 $F_V(0)$ & 0.0242  &0.0270   &  0.0262(5) &
                                             0.0258(17)~\cite{Bychkov:2008ws}
& $\hspace{0.05cm}$ 0.0271 & 0.0269 \\    
 $a_{V}$  &         & 0.0191  &  0.0332(42)&
                                             0.10(6)~\cite{Bychkov:2008ws}
& $\hspace{0.05cm}$ 0.0287 & 0.0280 \\  
 $F_A(0)$ & 0.0239  &0.0132   &  0.0106(36)&
                                             0.0117(17)~\cite{Bychkov:2008ws}
& $\hspace{0.05cm}$ 0.0132 & 0.0116 \\ 
 $a_{A}$  &         &0.012    &  0.0191(61)& $-$
& $\hspace{0.05cm}$ 0.0192 & 0.0193 \\ 
 $R_A(0)$ &         &         &            &
                                             $0.059^{+0.009}_{-0.008}$~\cite{Egli:1989vu}
& $\hspace{0.05cm}$ 0.0462 & 0.0459 \\ \hline \hline  
 \end{tabular}
\end{table}
In Table~\ref{tab:1}, we list the results for the form factors at
$Q^2=0$ and slope parameters that are defined as 
\begin{align}
F_{V}(Q^2)=\frac{F_{V}(0)}{1+ a_{V} \frac{Q^2}{m_{\pi}^2}},\;\;\;
F_{A}(Q^2)=\frac{F_{A}(0)}{1+ a_{A} \frac{Q^2}{m_{\pi}^2}},
\label{eq:SingPolFit}
\end{align} 
where $a_{V}$ and $a_A$ denote the slope parameter for the vector and
axial-vector form factors, respectively. 
The values of the vector and axial-vector form factors at $Q^2=0$ are 
respectively given as $F_V(0)=0.0271$, $F_A(0)=0.0132$, and
$R_A(0)=0.0462$ when the dynamical quark mass from the instanton
vacuum is used. The dipole-type parametrization yields
$F_V(0)=0.0269$, $F_A(0)=0.0116$, and $R_A(0)=0.0459$. As expected
from the results for the form factors shown in
Fig.~\ref{fig:3}, the values of $F_A(0)$ from these two
different forms of the dynamical quark mass are around $12\,\%$
different each other.  
The results are in good agreement with the experimental data 
$F_V^{\mathrm{exp}}=0.0258(17)$ and $F_A^{\mathrm{exp}} =
0.0117(17)$~\cite{Bychkov:2008ws}. It is also of 
great interest to compare the present results with those from other
works. Unterdorfer and Pichl~\cite{Unterdorfer:2008zz} analyzed the
vector and axial-vector form factors of pion radiative decay,
combining the results from $\chi$PT wirh a large $N_c$ expansion and
experimental data on other decays. The results are obtained as
$F_V(0)=0.0262(5)$ and $F_A(0)=0.0106(36)$, which are in good
agreement with the present results. Courtoy and
Noguera~\cite{Courtoy:2007vy} employed the NJL model to study the
$\pi$ photo-transition amplitude and derived from it the pion form
factors as $F_V(0)=0.0242$ and $F_A(0)=0.0239$. So, the result of the
vector form factor is comparable with that of the present work whereas
that of the axial-vector form factor is two times larger than this
one.  

The results of Ref.~\cite{GomezDumm:2012qh} are
especially interesting, since the nonlocal NJL model used in
Ref.~\cite{GomezDumm:2012qh} has several aspects in common with the
present model. In Ref.~\cite{GomezDumm:2012qh}, three different 
parameter sets were adopted, among which the results
with set A are compared with the present ones. Those from
Ref.~\cite{GomezDumm:2012qh} with set A are listed in
Table~\ref{tab:1} and are in good agreement with the present results
except for the slope parameters. It implies that the vector and
axial-vector form factors they obtained fall off more slowly than the
present ones. What is interesting is that the value $a_V=0.032$, which
is derived from the empirical fit to $\pi^0\to \gamma \gamma^*$
experimental data in Ref.~\cite{GomezDumm:2012qh}, is in good
agreement with that of the present work. 

In $\chi$PT, the values of $F_A(0)$ and $R_A(0)$ are given in terms of
the low-energy constants (LECs), $L_9$ and $L_{10}$
\begin{align}
F_A(0)=\frac{4\sqrt{2}m_\pi}{f_\pi}(L_9 + L_{10}),\;\;\;
R_A(0)=\frac{4\sqrt{2}m_\pi}{f_\pi}L_9.
\end{align}
We obtain the values from the present numerical calculation as Table \ref{tab:2}.
\begin{table}[htp]
\centering
\caption{The results of the low-energy constants. ``D.P.'' stands 
  for the results derive from the instanton vacuum, whereas ``Dipole''
  denotes those obtained from the dipole-type parametrization of the
  dynamical quark mass. } 
\label{tab:2}
\begin{tabular}{m{2.5cm}m{2.2cm}m{3.0cm}m{2.5cm}}\hline
       & $\hspace{0.5cm}L_9^r$    
       &  $\hspace{0.7cm}L_{10}^r$  
       &  $L_{9}^r+L_{10}^r$  \\ 
       \hline \hline
D.P.   & $5.43$   & $-3.88$ & $1.55$ \\
Dipole & $5.41$   & $-4.05$ & $1.36$ \\
\hline \hline    
\end{tabular}
\end{table}

It is also of interest to extract the parameters for the
parametrization of the vector and axial-vector form factors for $\pi$
radiative weak decays. In lattice QCD, the $p$-pole parametrization for a
form factor is often utilized~\cite{Brommel:2006ww, Brommel:2007xd},
which is different from the typical parametrization given in
Eq.(\ref{eq:SingPolFit}). Then, the present three transition form
factors can be parametrized as  
\begin{align}
F_{V}(Q^2)= \frac{F_{V}(0)}{\left(1+\frac{Q^2}{p_V m_{p_{V}}^2}
  \right)^{p_{V}}},\;\;\; 
F_{A}(Q^2)= \frac{F_{A}(0)}{\left(1+\frac{Q^2}{p_A
  m_{p_{A}}^2}\right)^{p_{A}}},\;\;\; 
R_{A}(Q^2)= \frac{R_{A}(0)}{\left(1+\frac{Q^2}{p_R
  m_{p_{R}}^2}\right)^{p_{R}}}, 
\end{align}
where the results of the corresponding parameters are listed in
Table~\ref{tab:3}. 
\begin{table}[htp]
\centering
\caption{The results of the p-pole parameters. ``D.P.'' stands 
  for the results derive from the instanton vacuum, whereas ``Dipole''
  denotes those obtained from the dipole-type parametrization of the
  dynamical quark mass. }
\label{tab:3}
\begin{tabular}{m{2cm}m{1cm}m{2cm}m{1cm}m{2cm}m{1cm}m{2cm}}\hline
       & $P_V$  & $M_{P_V}$  & $P_A$ & $M_{P_A}$ & $P_R$ & $M_{P_R}$ \\
       \hline \hline
D.P.   & $1.16$   & $0.843$ GeV & $1.48$ & $1.05$ GeV & $0.757$ & $1.10$ GeV \\
Dipole & $1.34$   & $0.870$ GeV & $1.59$ & $1.05$ GeV & $0.734$ & $1.12$ GeV \\
\hline \hline 
\end{tabular}
\end{table}

\section{Summary and conclusion}
In the present work, we aimed at investigating the form factors for
pion radiative weak decays, based on the gauged effective chiral
action derived from the instanton vaccuum. We computed the vector and
axial-vector transition form factors $F_V(Q^2)$, $F_A(Q^2)$, and
$R_A(Q^2)$, employing the momentum-dependent dynamical quark mass from
the instanton vacuum and that with the dipole-type
parametrization. The nonlocal contributions, which arise from the
gauging of the effective chiral action, enhance the vector form factor 
by about $20\,\%$, whereas they reduce the axial-vector form factor
$F_A(Q^2)$ by about 30 \%. The nonlocal terms influence the second
axial-vector form factor marginally. The difference between the
results from the instanton vacuum and those with the dipole-type
dynamical quark mass is almost the same except for the axial-vector
form factor for which the result with the dipole-type parametrization
is about $12\,\%$ smaller than that from the instanton vacuum. The
present results were compared with the experimental data and were
found to be in good agreement with the data except for the slope
parameter $a_V$. We also derived the low-energy constants $L_9^r$ and
$L_{10}^r$. Finally, we parametrized the form factors, using the
$p$-pole parametrization, which can be used to compare the present
results with those from the lattice data.  

It is also of interest to consider other types of the form factors for
pion radiative decay such as tensor transition form factors within the
present framework. These tensor form factors may give a clue about a
right direction beyond the Standard model. Moreover, they will provide
an opportunity to understand generalized transition form factors
related to the generalized parton distributions for weak
processes.  Another interesting
decay is kaon radiative decay, which will play a role of the
touchstone of understanding the effects of flavor SU(3) symmetry in
mesonic weak decays. The corresponding works are under way.
\section*{Acknowledgments}
We are grateful to A. Hosaka and H.D. Son for useful discussion. 
This work was supported by Inha University Research Grant. 

\end{document}